\documentclass{article}
\usepackage{spconf,amsmath,graphicx,url,times}
\usepackage{color}
\usepackage[nolist]{acronym}
\usepackage{mathtools}
\usepackage{cite}  
\usepackage{caption}
\usepackage{subcaption}
\usepackage{wrapfig}
\usepackage{environ}         
\usepackage{etoolbox}  
\usepackage{hyperref}
\usepackage[subtle]{savetrees}
\usepackage{float}
\usepackage{amssymb}
\usepackage{etoolbox}
\usepackage{glossaries}

\newcommand\norm[1]{\lVert#1\rVert}

\title{BINAURAL SPEECH ENHANCEMENT USING DEEP COMPLEX CONVOLUTIONAL TRANSFORMER NETWORKS}

 \name{Vikas Tokala$^{^1}$,
      Eric Grinstein$^{1}$,
      Mike Brookes$^{^1}$,
      Simon Doclo$^{^2}$,
      Jesper Jensen${^{3,4}}$,
      Patrick A. Naylor $^1$\sthanks{This work was supported by funding from the European Union’s Horizon 2020 research and innovation programme under the Marie Skłodowska-Curie grant agreement No 956369 and the UK Engineering and Physical Sciences Research Council [grant number EP/S035842/1]}} 
\address{$^{1}$Department of Electrical and Electronic Engineering, Imperial College London, UK\\
         $^{2}$ Department of Medical Physics and Acoustics, University of Oldenburg, Germany.\\
        $^{3}$  Demant A/S, Smørum, Denmark.\\
        $^{4}$ Department of Electronic Systems, Aalborg University, Denmark
}

\begin{acronym}
\acro{SNR}{Signal-to-Noise Ratio}
\acro{FwSSNR}{Frequency-weighted Segmental SNR}
\acro{SOBM}{STOI-optimal Binary Mask}
\acro{STOI}{Short-Time Objective Intelligibility}
\acro{WSTOI}{Weighted STOI}
\acro{HSWOBM}{High-resolution Stochastic WSTOI-optimal Binary Mask}
\acro{STFT}{Short Time Fourier Transform}
\acro{ILD}{Interaural Level Differences}
\acro{IPD}{Interaural Phase Differences}
\acro{ITD}{Interaural Time Differences}
\acro{LSA}{Log Spectral Amplitude}
\acro{OM-LSA}{Optimally-modified Log Spectral Amplitude}
\acro{DNN}{Deep Neural Network}
\acro{RNN}{Recurrent Neural Network}
\acro{ERB}{Equivalent Rectangular Bandwidth}
\acro{DFT}{Discrete Fourier Transform}
\acro{VSSNR}{Voiced-Speech-plus-Noise to Noise Ratio}
\acro{SPP}{Speech Presence Probability}
\acro{HRIRs}{Head Related Impulse Response}
\acro{MBSTOI}{Modified Binaural STOI}
\acro{ReLU}{Rectified Linear Unit}
\acro{SNR}{Signal-to-Noise Ratio}
\acro{ISTFT}{Inverse STFT}
\acro{TF}{Time-Frequency}
\acro{WGN}{White Gaussian Noise}
\acro{RTF}{Relative Transfer Function}
\acro{EVD}{Eigen Value Decomposition}
\acro{PSD}{Power Spectral Density}
\acro{IRM}{Ideal Ratio Mask}
\acro{NPM}{Normalized Projection Misalignment}
\acro{CRM}{Complex Ratio Mask}
\acro{FTB}{Frequency Transformation Block}
\acro{FTM}{Frequency Transformation Matrix}
\acro{FAL}{Frequency Attention Layer}
\acro{MWF}{Multichannel Wiener Filter}
\acro{MVDR}{Minimum Variance Distortionless Response}
\acro{MSC}{Magnitude Squared Coherence}
\acro{VAD}{Voice Activity Detector}
\acro{PESQ}{Preceptual Evalution of Speech Quality}
\acro{MLP}{Multi-Layer Perceptrons}
\acro{AED}{Auto Encoder-Decoder}
\acro{CED}{Convolutional Encoder-Decoder}
\acro{CNN}{Convolutional Neural Network}
\acro{CRN}{Convolutional Recurrrent Network}
\acro{LSTM}{Long short-term memory}
\acro{TNN}{Transformer Neural Networks}
\acro{NLP}{Natural Language Processing}
\acro{IBM}{Ideal Binary Mask}
\acro{IRM}{Ideal Ratio Mask}
\acro{GRU}{Gated Recurrent Unit}
\acro{HRIRs}{Head Related Impulse Response}
\acro{SSN}{Speech Shaped Noise}
\acro{HATS}{Head and Torso Simulator}
\acro{PReLU}{Parametric Rectified Linear Unit}
\acro{BSOBM}{Binaural STOI-Optimal Masking}
\acro{BMWF}{Binaural MWF}
\acro{MIMO}{Multiple Input Multiple Output}
\acro{SegSNR}{Segmental SNR}
\acro{RIR}{Room Impulse Responses}
\acro{BiTasNet}{Binaural TasNet}
\acro{SI-SNR}{Scale Invariant SNR}
\acro{BCCTN}{Binaural Complex Convolutional Transformer Network}
\acro{BRIR}{Binaural Room Impulse Responses}
\end{acronym}
\begin{document}
\ninept
\maketitle

\begin{sloppy}

\begin{abstract}
Studies have shown that in noisy acoustic environments, providing binaural signals to the user of an assistive listening device may improve speech intelligibility and spatial awareness. This paper presents a binaural speech enhancement method using a complex convolutional neural network with an encoder-decoder architecture and a complex multi-head attention transformer. The model is trained to estimate individual complex ratio masks in the time-frequency domain for the left and right-ear channels of binaural hearing devices. The model is trained using a novel loss function that incorporates the preservation of spatial information along with speech intelligibility improvement and noise reduction. Simulation results for acoustic scenarios with a single target speaker and isotropic noise of various types show that the proposed method improves the estimated binaural speech intelligibility and preserves the binaural cues better in comparison with several baseline algorithms.
\end{abstract}

\vspace{-0.5cm}
\section{Introduction}
\label{sec:intro}
 Binaural speech enhancement has been established in recent years as the state-of-the-art approach for enhancement in hearing aids and augmented/virtual reality devices \cite{Doclo2008, Guiraud2022}. Binaural signals contain the spatial characteristics of sounds, which carry the necessary information for accurate sound source localization \cite{hawley2004}. Moreover, binaural unmasking effects have been found to increase speech intelligibility therefore accentuating the importance of preservation of interaural cues for binaural signals along with noise reduction \cite{Beutelmann2006}. \ac{ILD}, and \ac{ITD} or \ac{IPD} are the primary cues helpful in localizing and boosting the perceived loudness of sounds, and improving speech intelligibility\cite{blauert1997}. Binaural speech enhancement using multichannel Wiener filters \cite{Hadad2015, Klasen2006}, beamforming \cite{Doclo2008}, and mask-based enhancement methods \cite{Tokala2022, Moore2018b} has been previously proposed. In \cite{Han2020}, a time domain \ac{CED} model for binaural speech separation was proposed and achieved state-of-the-art performance. In contrast to binaural methods, monaural speech enhancement approaches operating on each binaural channel independently enhance the signals but at the cost of damaging vital binaural cues.  Monaural speech enhancement methods using deep learning techniques have shown significant results in both the time domain \cite{Stoller2018a, Luo2018} and the \ac{TF} domain \cite{Tan2018, Yin2020a}.  

In the \ac{TF} domain, spectrograms are used as the input to the network \cite{Tokala2022, Yin2020a, Hu2020}. Most of the \ac{TF} domain methods rely only on magnitude-based enhancement, and the noisy \ac{TF} phase is used in the reconstruction of the enhanced speech signal \cite{Tokala2022,Kim2020a}. One of the ways to address the issue of optimal phase estimation for signal reconstruction is to jointly estimate the \ac{TF} phase and magnitude, which can be achieved by using complex-valued spectrograms. Monaural speech enhancement methods using complex-valued networks have shown promising results and have outperformed real-valued networks \cite{Kim2020a, Hu2020}. The \ac{CRN} introduced in \cite{Tan2018} employed a \acf{CED} architecture with \ac{LSTM} blocks placed in between the encoder and decoder. Moreover, Attention-based \ac{TNN} have shown state-of-the-art performance on \ac{NLP} problems compared to other \ac{DNN} models \cite{Vaswani2017}. Speech enhancement using attention models has been demonstrated in \cite{Kim2020a} with promising results.

In \cite{Hu2020}, a deep complex \ac{CRN} was trained to optimize the \ac{SI-SNR} for monaural speech signals. However, using a similar approach for binaural signals could be damaging to the interaural cues. More specifically, for the case of binaural signals, phase information is vital for preserving the \ac{IPD} values and the enhanced signals should retain level differences as the original signal to have the same \ac{ILD}. Even if the model achieves significant noise reduction and improves speech intelligibility, altering the level and phase information would modify the spatial information of the target and therefore compromise the localization and spatial awareness of the listener \cite{blauert1997, Beutelmann2006}. 

In this paper, we propose a method that uses a complex-valued \acf{CED} based transformer network which enables phase-aware training \cite{Hu2020, Trabelsi2018} for binaural speech and introduces terms in the loss function to simultaneously improve speech intelligibility and preserve the interaural cues of the speech signal. 



\section{Model Architecture}
\label{sec:method}
 \begin{figure*}
    \begin{subfigure}[b]{0.7\textwidth}
    \includegraphics[width=\textwidth]{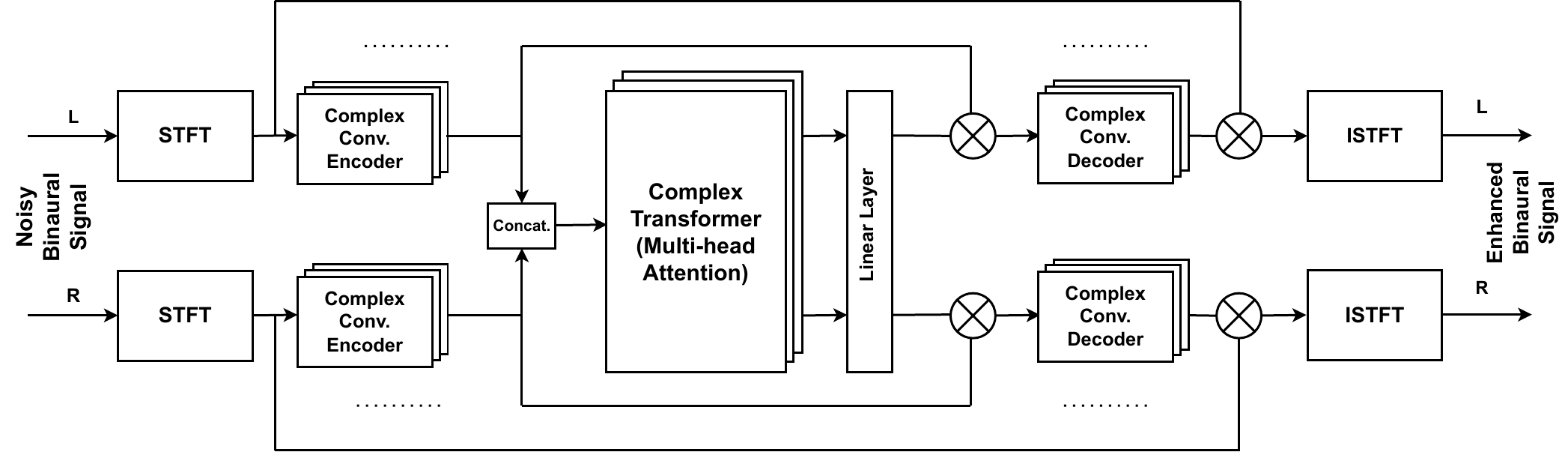}
    \caption{}
    \label{fig:BlockDiag}
      \end{subfigure}
        \begin{subfigure}[b]{0.3\textwidth}
    \includegraphics[scale=0.3]{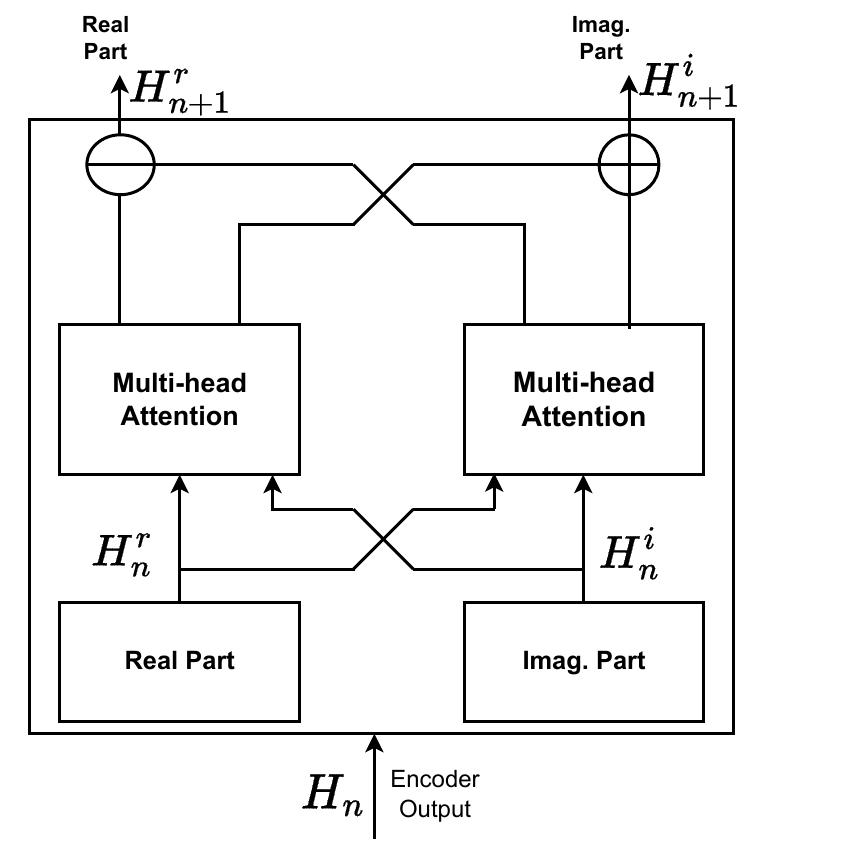}
    \caption{}
    \label{fig:BlockDiagtrans}
      \end{subfigure}
      \vspace{-.5cm}
    \caption{Architecture of (a) the proposed model and (b) the complex transformer block which implements \eqref{transEqn1} and \eqref{Attneqn1}.}
    \vspace{-0.5cm}
\end{figure*}

The proposed \ac{BCCTN} model uses a \acf{CED} structure with a transformer block between the encoder and decoder and is trained to estimate an individual \ac{CRM} for each channel. The block diagram of the architecture is shown in Fig.~\ref{fig:BlockDiag}. A \ac{CED} architecture for monaural speech enhancement has been previously introduced in \cite{Stoller2018a, Tan2018, Hu2020 }. The proposed \ac{MIMO} architecture uses a similar structure that has in this work newly modified to work with binaural signals by using individual encoder and decoder blocks for each channel. The \ac{STFT} blocks transform the signals into the \ac{TF} domain. The encoder block is made of 6 complex convolutional layers with \ac{PReLU} activation and employs batch normalization. The convolutional encoder blocks help in identifying the local patterns in the input spectrogram \cite{Stoller2018a, Hu2020}. Individual encoder blocks are used for the left and right-ear channels as the network needs to estimate two individual \ac{CRM}s. The encoded information from both channels is concatenated and supplied as the input to the transformer. The transformer block consists of multi-head attention layers based on the architecture proposed in \cite{Vaswani2017}. The structure of the complex transformer is shown in Fig.~\ref{fig:BlockDiagtrans}. The real and imaginary output of the transformer $H_{n+1}$ for the $(n+1)^{th}$ hidden state are given by
\vspace{-0.1cm}

\begin{equation}
\label{transEqn1}
    H^r_{n+1} =  ( H^r_n \circledast H^r_n  ) - ( H^i_n \circledast H^i_n  ),
\end{equation}
\vspace{-0.4cm}
\begin{equation}\label{Attneqn1}
     H^i_{n+1} =( H^r_n \circledast H^i_n ) + ( H^i_n \circledast H^r_n  ),
\end{equation}
 where $H^r_n$ and $H^i_n$ are the real and imaginary parts of the encoder output $H_n$. The multi-head attention operation is denoted by $\circledast$. The transformer block focuses on identifying relationships within the encoded information from both channels \cite{Vaswani2017,Kim2020a}. The convolutional decoder consists of 6 transposed complex convolutional blocks which are symmetric in design to the convolutional layers of the encoder to reconstruct the signal to its original size using the processed feature information from the transformer. Skip connections are placed between each encoder and decoder layer based on the \ac{CRN} architecture \cite{Tan2018} which concatenates the output of each encoder block to the decoder layer. This improves the information flow and facilitates network optimization \cite{Tan2018}. The left and right channel decoders output individual \ac{CRM}s that are applied to the noisy binaural signal for enhancement. The \ac{ISTFT} blocks in Fig.~\ref{fig:BlockDiag} transform the enhanced \ac{TF} domain signal back into the time domain. Implementation code is available online \footnote{\url{https://github.com/VikasTokala/BCCTN}}.

 \vspace{-.1cm}
\section{Signal Model and Loss Function}
\label{sec:Sig_model}
For the left channel, the noisy time-domain input signal $y_{L}$ is given by 
\begin{equation}
    y_{L}(t) = s_{L}(t) + v_{L}(t),
\end{equation}
where $s_{L}$ is the anechoic clean speech signal, $v_{L}$ is the noise and  $t$ is the discrete-time index. The \ac{STFT} is used to transform the signals into the \ac{TF} domain and the respective \ac{TF} representations are $Y_{L}(k,\ell)$, $S_{L}(k,\ell)$ and $V_{L}(k,\ell)$ with $k$ and $\ell$ being the frequency and time frame indices respectively. During training, the network learns to estimate a \ac{CRM}, $M_{L}(k,\ell)$ which is applied to the noisy signal $Y_{L}$ to obtain the enhanced speech signal $\hat{S}_{L}$ for the left ear. The right channel is described similarly with a $R$ subscript. For clarity, the $L$ and $R$ indices are omitted for the remainder of this paper. The enhanced speech is obtained for each channel by applying the estimated complex mask $\left({M}_r + j {M }_i \right)$ to the complex-valued noisy signal $\left({Y}_r + j {Y }_i\right)$ in the \ac{TF} domain (omitting $k$ and $\ell$ indices),
\begin{equation}
    \hat{S}_r + j \hat{S}_i = \left({M}_r + j {M }_i \right) \cdot \left({Y}_r + j {Y }_i\right),
\end{equation}
where $r$ and $i$ indicate the real and imaginary parts. The computed \ac{CRM} \cite{Williamson2015} is given by 
\begin{equation}
    {M}_r + j {M }_i = \frac{\hat{S}_r + j \hat{S}_i}{{Y}_r + j {Y }_i} = \frac{{Y}_r \hat{S}_r+{Y}_i \hat{S}_i}{Y_r^2 + Y_i^2} + j \frac{{Y}_r \hat{S}_i-{Y}_i \hat{S}_r}{Y_r^2 + Y_i^2}.
\end{equation}
\vspace{-.5cm}
\subsection{Loss Function}
\label{sec:LossFunc}
The proposed loss function for model training contains four terms and optimizes the network for noise reduction, intelligibility improvement, and interaural cue preservation. The proposed loss function $\mathcal{L}$ is given by
\vspace{-.1cm}
\begin{equation} \label{lossfunc}
\mathcal{L} = \alpha \mathcal{L}_{SNR} + \beta \mathcal{L}_{STOI} +   \gamma \mathcal{L}_{ILD} + \kappa \mathcal{L}_{IPD},
\end{equation}
where $\mathcal{L}_{SNR}$ is the \ac{SNR} loss, $\mathcal{L}_{STOI}$ is the \ac{STOI} \cite{Taal2010} loss, and $\mathcal{L}_{ILD}$ and $\mathcal{L}_{IPD}$ are the proposed \ac{ILD} and \ac{IPD} error losses which are functions of both $\hat{S}_L$ and $\hat{S}_R$. The parameters $\alpha$, $\beta$, $\gamma$, and $\kappa$ are the weights applied to each term. 

The \ac{SNR} of the enhanced signal, $\hat{\mathbf{s}}$, is defined as
\begin{equation} \label{snr}
   \ac{SNR}( \mathbf{s},\hat{\mathbf{s}}) =  10 \log_{10} \left( \frac{\norm{\mathbf{s}}^2}{\norm{\mathbf{e}_{noise}}^2}\right),
    \end{equation}
where $\mathbf{e}_{noise} = \hat{\mathbf{s}} - \mathbf{s}$ with $\mathbf{s}$ and $\hat{\mathbf{s}}$ being the clean and enhanced signal vectors respectively and $\norm{.}$ is the L2 norm. We define $\mathcal{L}_{SNR}$ to be the mean of the left and right-ear channel values and append a negative sign to maximize the \ac{SNR} value, such that $\mathcal{L}_{SNR} = - \left(\ac{SNR}_{L} + \ac{SNR}_{R} \right)/2$. 

While $\mathcal{L}_{SNR}$ optimizes the network for noise reduction, $\mathcal{L}_{STOI}$ is designed for intelligibility improvement. Similar to $\mathcal{L}_{SNR}$ we optimize the network to maximize intelligibility and $\mathcal{L}_{STOI}$ \cite{Taal2010} is computed for the left and right channels individually and averaged so that $\mathcal{L}_{STOI}~=~ - \left(\ac{STOI}_{L} + \ac{STOI}_{R} \right) /2$ \cite{Manuel2023}. 

As the network is trained to compute two individual \ac{CRM}s for binaural speech, it has to be forced to preserve the interaural cues of the target speech while enhancing the noisy signal. To optimize the network for cue preservation, \ac{ILD} and \ac{IPD} errors of the target speech are computed for the enhanced speech signal. The \ac{ILD} and \ac{IPD} for the clean speech signal are given by

\begin{equation} \label{ild}
    ILD_S(k,\ell) = 20 \log_{10} \left ( \frac{\vert S_{L}(k,\ell) \vert }{ \vert S_{R}(k,\ell) \vert}\right),
\end{equation}
\begin{equation}
\label{ipd}
    IPD_S (k, \ell) = \arctan \left ( \frac{S_{L}(k,\ell) }{S_{R}(k,\ell)} \right).
\end{equation}

\noindent The \ac{ILD} and \ac{IPD} for the enhanced speech are calculated similarly to \eqref{ild} and \eqref{ipd}. The $\mathcal{L}_{ILD}$ and $\mathcal{L}_{IPD}$ terms are given by,
\vspace{-.2cm}
\begin{equation} \label{ildloss}
  \mathcal{L}_{ILD} = \frac{1}{N} \sum_{k,\ell} \mathcal{M}(k,\ell)  \left(  \vert ILD_{S} (k,\ell) - ILD_{\hat{S}} (k,\ell) \vert \right ),
\end{equation}

\begin{equation}\label{ipdloss}
    \mathcal{L}_{IPD} = \frac{1}{N} \sum_{k,\ell} \mathcal{M}(k,\ell) \vert IPD_{S} (k,\ell) - IPD_{\hat{S}} (k,\ell) \vert
\end{equation}

\noindent where $N = \sum_{k,\ell} \mathcal{M}(k,\ell) $ is the total number of speech-active frequency and time bins determined from the mask. To compute the \ac{ILD} and \ac{IPD} errors only in the speech-active regions, an \ac{IBM} \cite{Wang2005b} $\mathcal{M}$ is computed by choosing the \ac{TF} bins which have energy above a threshold. The energy $E(k,\ell)$ of the clean signal is given by
\begin{equation}
    E(k,\ell) = 10 \log_{10} {\vert {S}(k,\ell) \vert}^2.
\end{equation}
 The \ac{IBM} $\mathcal{M} (k,\ell)$ that defines the speech active \ac{TF} tiles is then defined as,
\vspace{-0.15cm}
\begin{equation}
    \mathcal{M}(k,\ell) = \begin{dcases}
    1 & E(k,\ell) > \max_{\ell} \left ( E(k,\ell) \right) - \mathcal{T} \\
    0 & \mathrm{otherwise}.
    \end{dcases}
\end{equation}
$\max_l \left ( E(k,\ell) \right) $ is the maximum energy computed for each frequency $k$. Individual \ac{IBM}s, $\mathcal{M}_L$ and $\mathcal{M}_R$ are computed for the left and right-ear channels. The final mask $\mathcal{M}$ is obtained by choosing the bins that have energy above the threshold, $\max_{\ell} \left ( E(k,\ell) \right) - \mathcal{T}$, in both channels and is given by

\begin{equation}
    \mathcal{M}(k,\ell) = \mathcal{M}_L(k,\ell) \odot \mathcal{M}_R (k,\ell),
\end{equation}
\begin{figure}
    \centering
    \includegraphics[width=0.5\textwidth]{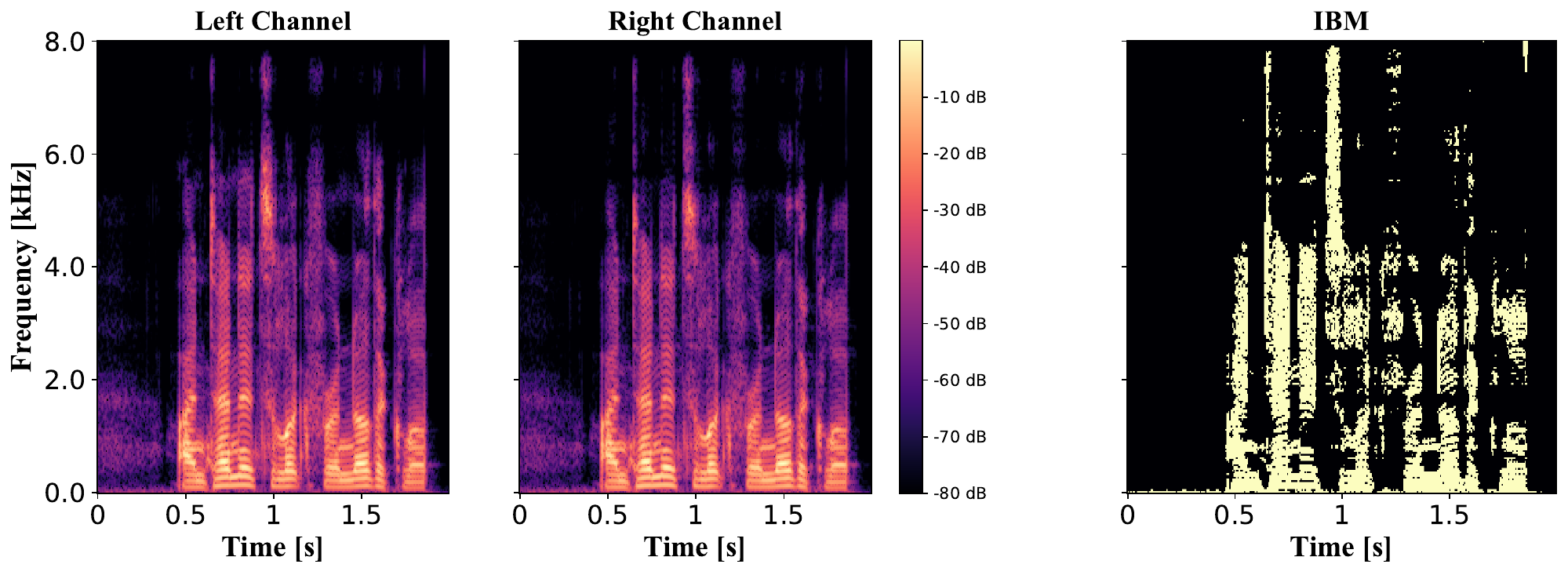}
    \vspace{-0.4cm}
    \caption{Spectrograms of the left and right-ear clean speech signals and the corresponding \ac{IBM} computed for interaural cue error masking.}
    \label{fig:ild_ipd_IBM}
    \vspace{-.5cm}
\end{figure}
\noindent where $\odot$ denotes the Hadamard product. For training and evaluation, $\mathcal{T}=20$~dB was used \cite{Wang2005b}. As an example, Figure~\ref{fig:ild_ipd_IBM} shows the spectrograms of the clean speech signal and the corresponding target speech-based binary mask. Using the target speech-based mask guides the optimization of the network to focus on the preservation of the interaural cues of the target speech. The \ac{ILD} and \ac{IPD} errors are computed in the \ac{TF} domain and the \ac{SNR} and \ac{STOI} losses are computed in the time domain by synthesizing the waveform using the \ac{ISTFT}.


\begin{table*}[hbt!]
\Large
\centering
\resizebox{\textwidth}{!}{%
\begin{tabular}{|l|cccc|cccc|cccc|cccc|}
\hline
\multicolumn{1}{|c|}{\textbf{Input   SNR}} &
  \multicolumn{4}{c|}{\textbf{-6 dB}} &
  \multicolumn{4}{c|}{\textbf{-3 dB}} &
  \multicolumn{4}{c|}{\textbf{0 dB}} &
  \multicolumn{4}{c|}{\textbf{3 dB}} \\ \hline
\multicolumn{1}{|c|}{\textbf{Method}} &
  \multicolumn{1}{c|}{\textbf{MBSTOI}} &
  \multicolumn{1}{c|}{\textbf{$\Delta$SegSNR}} &
  \multicolumn{1}{c|}{\textbf{$\mathcal{L}_{ILD}$}} &
  \textbf{$\mathcal{L}_{IPD}$ } &
  \multicolumn{1}{c|}{\textbf{MBSTOI}} &
  \multicolumn{1}{c|}{\textbf{$\Delta$SegSNR}} &
  \multicolumn{1}{c|}{\textbf{$\mathcal{L}_{ILD}$}} &
  \textbf{$\mathcal{L}_{IPD}$ } &
  \multicolumn{1}{c|}{\textbf{MBSTOI}} &
  \multicolumn{1}{c|}{\textbf{$\Delta$SegSNR}} &
  \multicolumn{1}{c|}{\textbf{$\mathcal{L}_{ILD}$}} &
  \textbf{$\mathcal{L}_{IPD}$ } &
  \multicolumn{1}{c|}{\textbf{MBSTOI}} &
  \multicolumn{1}{c|}{\textbf{$\Delta$SegSNR}} &
  \multicolumn{1}{c|}{\textbf{$\mathcal{L}_{ILD}$}} &
  \textbf{$\mathcal{L}_{IPD}$ } \\ \hline \hline
\textbf{Noisy signal} &
  \multicolumn{1}{c|}{0.61} &
  \multicolumn{1}{c|}{0} &
  \multicolumn{1}{c|}{-} &
  - &
  \multicolumn{1}{c|}{0.69} &
  \multicolumn{1}{c|}{0} &
  \multicolumn{1}{c|}{-} &
  - &
  \multicolumn{1}{c|}{0.78} &
  \multicolumn{1}{c|}{0} &
  \multicolumn{1}{c|}{-} &
  - &
  \multicolumn{1}{c|}{0.8} &
  \multicolumn{1}{c|}{0} &
  \multicolumn{1}{c|}{-} &
  - \\ \hline
\textbf{BSOBM} \cite{Tokala2022} &
  \multicolumn{1}{c|}{0.63} &
  \multicolumn{1}{c|}{4.3} &
  \multicolumn{1}{c|}{0.94} &
  11 &
  \multicolumn{1}{c|}{0.7} &
  \multicolumn{1}{c|}{6.8} &
  \multicolumn{1}{c|}{1.27} &
  10 &
  \multicolumn{1}{c|}{0.76} &
  \multicolumn{1}{c|}{6.5} &
  \multicolumn{1}{c|}{1.05} &
  11 &
  \multicolumn{1}{c|}{0.78} &
  \multicolumn{1}{c|}{6.9} &
  \multicolumn{1}{c|}{1.08} &
  13 \\ \hline
\textbf{BiTasNet} \cite{Han2020} &
  \multicolumn{1}{c|}{0.69} &
  \multicolumn{1}{c|}{\textbf{14.5}} &
  \multicolumn{1}{c|}{0.86} &
  12 &
  \multicolumn{1}{c|}{0.76} &
  \multicolumn{1}{c|}{\textbf{13.1}} &
  \multicolumn{1}{c|}{0.79} &
  9 &
  \multicolumn{1}{c|}{0.82} &
  \multicolumn{1}{c|}{\textbf{12.8}} &
  \multicolumn{1}{c|}{0.74} &
  10 &
  \multicolumn{1}{c|}{0.86} &
  \multicolumn{1}{c|}{\textbf{11.6}} &
  \multicolumn{1}{c|}{0.67} &
  9 \\ \hline
\textbf{BCCTN-SNR} \eqref{snr} &
  \multicolumn{1}{c|}{0.63} &
  \multicolumn{1}{c|}{13.2} &
  \multicolumn{1}{c|}{0.74} &
  11 &
  \multicolumn{1}{c|}{0.71} &
  \multicolumn{1}{c|}{11.9} &
  \multicolumn{1}{c|}{0.95} &
  11 &
  \multicolumn{1}{c|}{0.77} &
  \multicolumn{1}{c|}{12.1} &
  \multicolumn{1}{c|}{0.6} &
  10 &
  \multicolumn{1}{c|}{0.83} &
  \multicolumn{1}{c|}{11} &
  \multicolumn{1}{c|}{0.86} &
  11 \\ \hline
\textbf{BCCTN-Proposed Loss} \eqref{lossfunc} &
  \multicolumn{1}{c|}{\textbf{0.73}} &
  \multicolumn{1}{c|}{14.3} &
  \multicolumn{1}{c|}{\textbf{0.61}} &
  \textbf{8} &
  \multicolumn{1}{c|}{\textbf{0.79}} &
  \multicolumn{1}{c|}{12.7} &
  \multicolumn{1}{c|}{\textbf{0.62}} &
  \textbf{7} &
  \multicolumn{1}{c|}{\textbf{0.85}} &
  \multicolumn{1}{c|}{12.7} &
  \multicolumn{1}{c|}{\textbf{0.4}} &
  \textbf{5} &
  \multicolumn{1}{c|}{\textbf{0.87}} &
  \multicolumn{1}{c|}{11.5} &
  \multicolumn{1}{c|}{\textbf{0.36}} &
  \textbf{4} \\ \hline
\end{tabular}%
}
\Large
\centering
\resizebox{\textwidth}{!}{%
\begin{tabular}{|l|cccc|cccc|cccc|cccc|}
\hline
\multicolumn{1}{|c|}{\textbf{Input   SNR}} &
  \multicolumn{4}{c|}{\textbf{6 dB}} &
  \multicolumn{4}{c|}{\textbf{9 dB}} &
  \multicolumn{4}{c|}{\textbf{12 dB}} &
  \multicolumn{4}{c|}{\textbf{15 dB}} \\ \hline
\multicolumn{1}{|c|}{\textbf{Method}} &
  \multicolumn{1}{c|}{\textbf{MBSTOI}} &
  \multicolumn{1}{c|}{\textbf{$\Delta$SegSNR}} &
  \multicolumn{1}{c|}{\textbf{$\mathcal{L}_{ILD}$}} &
  \textbf{$\mathcal{L}_{IPD}$ } &
  \multicolumn{1}{c|}{\textbf{MBSTOI}} &
  \multicolumn{1}{c|}{\textbf{$\Delta$SegSNR}} &
  \multicolumn{1}{c|}{\textbf{$\mathcal{L}_{ILD}$}} &
  \textbf{$\mathcal{L}_{IPD}$ } &
  \multicolumn{1}{c|}{\textbf{MBSTOI}} &
  \multicolumn{1}{c|}{\textbf{$\Delta$SegSNR}} &
  \multicolumn{1}{c|}{\textbf{$\mathcal{L}_{ILD}$}} &
  \textbf{$\mathcal{L}_{IPD}$ } &
  \multicolumn{1}{c|}{\textbf{MBSTOI}} &
  \multicolumn{1}{c|}{\textbf{$\Delta$SegSNR}} &
  \multicolumn{1}{c|}{\textbf{$\mathcal{L}_{ILD}$}} &
  \textbf{$\mathcal{L}_{IPD}$ } \\ \hline \hline
\textbf{Noisy signal} &
  \multicolumn{1}{c|}{0.88} &
  \multicolumn{1}{c|}{0} &
  \multicolumn{1}{c|}{-} &
  - &
  \multicolumn{1}{c|}{0.92} &
  \multicolumn{1}{c|}{0} &
  \multicolumn{1}{c|}{-} &
  - &
  \multicolumn{1}{c|}{0.95} &
  \multicolumn{1}{c|}{0} &
  \multicolumn{1}{c|}{-} &
   &
  \multicolumn{1}{c|}{0.95} &
  \multicolumn{1}{c|}{-} &
  \multicolumn{1}{c|}{-} &
  - \\ \hline
\textbf{BSOBM} \cite{Tokala2022} &
  \multicolumn{1}{c|}{0.82} &
  \multicolumn{1}{c|}{5.6} &
  \multicolumn{1}{c|}{1.14} &
  13 &
  \multicolumn{1}{c|}{0.84} &
  \multicolumn{1}{c|}{3.6} &
  \multicolumn{1}{c|}{1.03} &
  12 &
  \multicolumn{1}{c|}{0.84} &
  \multicolumn{1}{c|}{1.3} &
  \multicolumn{1}{c|}{1.6} &
  12 &
  \multicolumn{1}{c|}{0.82} &
  \multicolumn{1}{c|}{-1.1} &
  \multicolumn{1}{c|}{1.54} &
  8 \\ \hline
\textbf{BiTasNet}\cite{Han2020} &
  \multicolumn{1}{c|}{0.89} &
  \multicolumn{1}{c|}{\textbf{9.9}} &
  \multicolumn{1}{c|}{0.63} &
  7 &
  \multicolumn{1}{c|}{0.92} &
  \multicolumn{1}{c|}{\textbf{8.6}} &
  \multicolumn{1}{c|}{0.55} &
  6 &
  \multicolumn{1}{c|}{0.93} &
  \multicolumn{1}{c|}{\textbf{7.2}} &
  \multicolumn{1}{c|}{0.35} &
  8 &
  \multicolumn{1}{c|}{0.93} &
  \multicolumn{1}{c|}{\textbf{5.6}} &
  \multicolumn{1}{c|}{0.46} &
  7 \\ \hline
\textbf{BCCTN-SNR} \eqref{snr} &
  \multicolumn{1}{c|}{0.87} &
  \multicolumn{1}{c|}{9.2} &
  \multicolumn{1}{c|}{0.82} &
  11 &
  \multicolumn{1}{c|}{0.9} &
  \multicolumn{1}{c|}{7.6} &
  \multicolumn{1}{c|}{0.66} &
  8 &
  \multicolumn{1}{c|}{0.93} &
  \multicolumn{1}{c|}{6.3} &
  \multicolumn{1}{c|}{0.8} &
  9 &
  \multicolumn{1}{c|}{0.92} &
  \multicolumn{1}{c|}{4.8} &
  \multicolumn{1}{c|}{0.87} &
  8 \\ \hline
\textbf{BCCTN-Proposed Loss} \eqref{lossfunc} &
  \multicolumn{1}{c|}{\textbf{0.91}} &
  \multicolumn{1}{c|}{9.7} &
  \multicolumn{1}{c|}{\textbf{0.34}} &
  \textbf{3} &
  \multicolumn{1}{c|}{\textbf{0.94}} &
  \multicolumn{1}{c|}{8.4} &
  \multicolumn{1}{c|}{\textbf{0.2}} &
 \textbf{2} &
  \multicolumn{1}{c|}{\textbf{0.96}} &
  \multicolumn{1}{c|}{7} &
  \multicolumn{1}{c|}{\textbf{0.19}} &
  \textbf{2} &
  \multicolumn{1}{c|}{\textbf{0.96}} &
  \multicolumn{1}{c|}{5.4} &
  \multicolumn{1}{c|}{\textbf{0.19}} &
  \textbf{2} \\ \hline
\end{tabular}%
}
\caption{Results for anechoic speech signals with isotropic noise averaged over all frames, frequency bins and utterances. $\Delta$ SegSNR \cite{Brookes1997} and  $\mathcal{L}_{ILD}$ \eqref{ildloss} are in dB, $\mathcal{L}_{IPD}$ \eqref{ipdloss} are in degrees.}
\label{tab_anechoic}
 \vspace{-0.3cm}
\end{table*}

\begin{table*}[hbt!]
\centering
\Large
\resizebox{\textwidth}{!}{%
\begin{tabular}{|l|cccc|cccc|cccc|cccc|}
\hline
\multicolumn{1}{|c|}{\textbf{Input   SNR}} &
  \multicolumn{4}{c|}{\textbf{-6 dB}} &
  \multicolumn{4}{c|}{\textbf{-3 dB}} &
  \multicolumn{4}{c|}{\textbf{0 dB}} &
  \multicolumn{4}{c|}{\textbf{3 dB}} \\ \hline
\multicolumn{1}{|c|}{\textbf{Method}} &
  \multicolumn{1}{c|}{\textbf{MBSTOI}} &
  \multicolumn{1}{c|}{\textbf{$\Delta$SegSNR}} &
  \multicolumn{1}{c|}{\textbf{$\mathcal{L}_{ILD}$}} &
  \textbf{$\mathcal{L}_{IPD}$ } &
  \multicolumn{1}{c|}{\textbf{MBSTOI}} &
  \multicolumn{1}{c|}{\textbf{$\Delta$SegSNR}} &
  \multicolumn{1}{c|}{\textbf{$\mathcal{L}_{ILD}$} }&
  \textbf{$\mathcal{L}_{IPD}$ } &
  \multicolumn{1}{c|}{\textbf{MBSTOI}} &
  \multicolumn{1}{c|}{\textbf{$\Delta$SegSNR}} &
  \multicolumn{1}{c|}{\textbf{$\mathcal{L}_{ILD}$}} &
  \textbf{$\mathcal{L}_{IPD}$ } &
  \multicolumn{1}{c|}{\textbf{MBSTOI}} &
  \multicolumn{1}{c|}{\textbf{$\Delta$SegSNR}} &
  \multicolumn{1}{c|}{\textbf{$\mathcal{L}_{ILD}$}} &
  \textbf{$\mathcal{L}_{IPD}$ } \\ \hline \hline
\textbf{Noisy signal} &
  \multicolumn{1}{c|}{0.59} &
  \multicolumn{1}{c|}{0} &
  \multicolumn{1}{c|}{-} &
  - &
  \multicolumn{1}{c|}{0.68} &
  \multicolumn{1}{c|}{0} &
  \multicolumn{1}{c|}{-} &
  - &
  \multicolumn{1}{c|}{0.76} &
  \multicolumn{1}{c|}{0} &
  \multicolumn{1}{c|}{-} &
  - &
  \multicolumn{1}{c|}{0.79} &
  \multicolumn{1}{c|}{0} &
  \multicolumn{1}{c|}{-} &
  - \\ \hline
\textbf{BSOBM}\cite{Tokala2022} &
  \multicolumn{1}{c|}{0.62} &
  \multicolumn{1}{c|}{2.9} &
  \multicolumn{1}{c|}{1.27} &
  17 &
  \multicolumn{1}{c|}{0.69} &
  \multicolumn{1}{c|}{4.8} &
  \multicolumn{1}{c|}{1.45} &
  18 &
  \multicolumn{1}{c|}{0.76} &
  \multicolumn{1}{c|}{3.9} &
  \multicolumn{1}{c|}{1.25} &
  13 &
  \multicolumn{1}{c|}{0.77} &
  \multicolumn{1}{c|}{3.8} &
  \multicolumn{1}{c|}{1.22} &
  12 \\ \hline
\textbf{BiTasNet} \cite{Han2020} &
  \multicolumn{1}{c|}{0.58} &
  \multicolumn{1}{c|}{10.1} &
  \multicolumn{1}{c|}{1.1} &
  14 &
  \multicolumn{1}{c|}{0.66} &
  \multicolumn{1}{c|}{\textbf{9.2}} &
  \multicolumn{1}{c|}{0.97} &
  12 &
  \multicolumn{1}{c|}{0.74} &
  \multicolumn{1}{c|}{\textbf{8.8}} &
  \multicolumn{1}{c|}{0.92} &
  11 &
  \multicolumn{1}{c|}{0.78} &
  \multicolumn{1}{c|}{\textbf{7.7}} &
  \multicolumn{1}{c|}{0.87} &
  10 \\ \hline
\textbf{BCCTN-SNR} \eqref{snr} &
  \multicolumn{1}{c|}{0.43} &
  \multicolumn{1}{c|}{9.6} &
  \multicolumn{1}{c|}{1.43} &
  16 &
  \multicolumn{1}{c|}{0.52} &
  \multicolumn{1}{c|}{8.9} &
  \multicolumn{1}{c|}{1.18} &
  13 &
  \multicolumn{1}{c|}{0.57} &
  \multicolumn{1}{c|}{8.2} &
  \multicolumn{1}{c|}{0.91} &
  12 &
  \multicolumn{1}{c|}{0.62} &
  \multicolumn{1}{c|}{6.8} &
  \multicolumn{1}{c|}{0.9} &
  11 \\ \hline
\textbf{BCCTN-Proposed Loss} \eqref{lossfunc} &
  \multicolumn{1}{c|}{\textbf{0.66}} &
  \multicolumn{1}{c|}{\textbf{10.3}} &
  \multicolumn{1}{c|}{\textbf{1.12}} &
  \textbf{12} &
  \multicolumn{1}{c|}{\textbf{0.74}} &
  \multicolumn{1}{c|}{9} &
  \multicolumn{1}{c|}{\textbf{0.72}} &
  \textbf{10} &
  \multicolumn{1}{c|}{\textbf{0.8}} &
  \multicolumn{1}{c|}{8.4} &
  \multicolumn{1}{c|}{\textbf{0.62}} &
  \textbf{8} &
  \multicolumn{1}{c|}{\textbf{0.83}} &
  \multicolumn{1}{c|}{7.1} &
  \multicolumn{1}{c|}{\textbf{0.45}} &
  \textbf{5} \\ \hline
\end{tabular}%
}
\Large
\centering
\resizebox{\textwidth}{!}{%
\begin{tabular}{|l|cccc|cccc|cccc|cccc|}
\hline
\multicolumn{1}{|c|}{\textbf{Input   SNR}} &
  \multicolumn{4}{c|}{\textbf{6 dB}} &
  \multicolumn{4}{c|}{\textbf{9 dB}} &
  \multicolumn{4}{c|}{\textbf{12 dB}} &
  \multicolumn{4}{c|}{\textbf{15 dB}} \\ \hline
\multicolumn{1}{|c|}{\textbf{Method}} &
  \multicolumn{1}{c|}{\textbf{MBSTOI}} &
  \multicolumn{1}{c|}{\textbf{$\Delta$SegSNR}} &
  \multicolumn{1}{c|}{\textbf{$\mathcal{L}_{ILD}$}} &
  \textbf{$\mathcal{L}_{IPD}$ } &
  \multicolumn{1}{c|}{\textbf{MBSTOI}} &
  \multicolumn{1}{c|}{\textbf{$\Delta$SegSNR}} &
  \multicolumn{1}{c|}{\textbf{$\mathcal{L}_{ILD}$}} &
  \textbf{$\mathcal{L}_{IPD}$ } &
  \multicolumn{1}{c|}{\textbf{MBSTOI}} &
  \multicolumn{1}{c|}{\textbf{$\Delta$SegSNR}} &
  \multicolumn{1}{c|}{\textbf{$\mathcal{L}_{ILD}$}} &
  \textbf{$\mathcal{L}_{IPD}$ } &
  \multicolumn{1}{c|}{\textbf{MBSTOI}} &
  \multicolumn{1}{c|}{\textbf{$\Delta$SegSNR}} &
  \multicolumn{1}{c|}{\textbf{$\mathcal{L}_{ILD}$}} &
  \textbf{$\mathcal{L}_{IPD}$ } \\ \hline \hline
\textbf{Noisy signal} &
  \multicolumn{1}{c|}{0.87} &
  \multicolumn{1}{c|}{0} &
  \multicolumn{1}{c|}{-} &
  - &
  \multicolumn{1}{c|}{0.92} &
  \multicolumn{1}{c|}{0} &
  \multicolumn{1}{c|}{-} &
  - &
  \multicolumn{1}{c|}{0.95} &
  \multicolumn{1}{c|}{0} &
  \multicolumn{1}{c|}{-} &
  - &
  \multicolumn{1}{c|}{0.95} &
  \multicolumn{1}{c|}{0} &
  \multicolumn{1}{c|}{-} &
  - \\ \hline
\textbf{BSOBM}\cite{Tokala2022} &
  \multicolumn{1}{c|}{0.81} &
  \multicolumn{1}{c|}{2.6} &
  \multicolumn{1}{c|}{1.19} &
  12 &
  \multicolumn{1}{c|}{0.84} &
  \multicolumn{1}{c|}{0.2} &
  \multicolumn{1}{c|}{1.7} &
  16 &
  \multicolumn{1}{c|}{0.85} &
  \multicolumn{1}{c|}{-2.2} &
  \multicolumn{1}{c|}{1.9} &
  12 &
  \multicolumn{1}{c|}{0.83} &
  \multicolumn{1}{c|}{-4.8} &
  \multicolumn{1}{c|}{2.1} &
  13 \\ \hline
\textbf{BiTasNet} \cite{Han2020} &
  \multicolumn{1}{c|}{0.85} &
  \multicolumn{1}{c|}{\textbf{6.9}} &
  \multicolumn{1}{c|}{0.81} &
  8 &
  \multicolumn{1}{c|}{0.9} &
  \multicolumn{1}{c|}{\textbf{5.8}} &
  \multicolumn{1}{c|}{0.59} &
  7 &
  \multicolumn{1}{c|}{0.93} &
  \multicolumn{1}{c|}{\textbf{4.8}} &
  \multicolumn{1}{c|}{0.53} &
  8 &
  \multicolumn{1}{c|}{0.92} &
  \multicolumn{1}{c|}{\textbf{3.7}} &
  \multicolumn{1}{c|}{0.59} &
  7 \\ \hline
\textbf{BCCTN-SNR} \eqref{snr} &
  \multicolumn{1}{c|}{0.73} &
  \multicolumn{1}{c|}{5.5} &
  \multicolumn{1}{c|}{1.26} &
  12 &
  \multicolumn{1}{c|}{0.81} &
  \multicolumn{1}{c|}{4.2} &
  \multicolumn{1}{c|}{0.88} &
  11 &
  \multicolumn{1}{c|}{0.9} &
  \multicolumn{1}{c|}{2.8} &
  \multicolumn{1}{c|}{0.73} &
  10 &
  \multicolumn{1}{c|}{0.88} &
  \multicolumn{1}{c|}{1.5} &
  \multicolumn{1}{c|}{0.91} &
  9 \\ \hline
\textbf{BCCTN-Proposed Loss} \eqref{lossfunc} &
  \multicolumn{1}{c|}{\textbf{0.89}} &
  \multicolumn{1}{c|}{6.1} &
  \multicolumn{1}{c|}{\textbf{0.38}} &
  \textbf{5} &
  \multicolumn{1}{c|}{\textbf{0.93}} &
  \multicolumn{1}{c|}{5} &
  \multicolumn{1}{c|}{\textbf{0.29}} &
  \textbf{3} &
  \multicolumn{1}{c|}{\textbf{0.96}} &
  \multicolumn{1}{c|}{4.6} &
  \multicolumn{1}{c|}{\textbf{0.21}} &
  \textbf{3} &
  \multicolumn{1}{c|}{\textbf{0.96}} &
  \multicolumn{1}{c|}{3.3} &
  \multicolumn{1}{c|}{\textbf{0.2}} &
  \textbf{2} \\ \hline
\end{tabular}%
}
\caption{Results for reverberant speech signals with isotropic noise and are averaged over all frames, frequency bins and utterances. $\Delta$ SegSNR \cite{Brookes1997} and $\mathcal{L}_{ILD}$ \eqref{ildloss} are in dB, $\mathcal{L}_{IPD}$ \eqref{ipdloss} are in degrees.}
\vspace{-0.55cm}
\label{tab_reverb}
\end{table*}

\section{Experiments}
\label{sec:Experiments}
\subsection{Datasets}
To generate binaural speech data, monaural clean speech signals were taken from the CSTR VCTK corpus \cite{Yamagishi2019} and were spatialized using the measured \ac{HRIRs} from \cite{Kayser2009}. The speech corpus \cite{Yamagishi2019} has around 13 hours of speech data uttered by 110 English speakers with various accents that were used to generate 2-second speech utterances and spatialized to have the left and right-ear channels. The dataset was made of 20000 speech utterances which were split into training, validation, and testing sets. Unseen speech data from the TIMIT corpus \cite{Garofolo1993} were also used for testing. Noise signals from the NOISEX-92 database \cite{Varga1993} were used to generate diffuse isotropic noise. Isotropic noise was generated using uncorrelated noise sources uniformly spaced every $5^\circ$ in the azimuthal plane\cite{Moore2018b} using HRIRs from \cite{Kayser2009}. Binaural signals were generated with the target speech placed at a random azimuth in the frontal plane ($-90 ^\circ$ to $+90^\circ$), using the \ac{HRIRs} from \cite{Kayser2009} recorded using a \ac{HATS}. For training, isotropic noise was added to the VCTK corpus\cite{Yamagishi2019} so that $(SNR_L + SNR_R)/2$ lies between -7~dB and 16~dB. The noise types used for training are \ac{WGN}, \ac{SSN}, factory noise, and office noise and, for evaluation, an additional car engine noise was included. The datasets were generated in the anechoic condition for training. The evaluation set consists of speech signals from the VCTK corpus\cite{Yamagishi2019} (i.e, ``matched" condition) and the TIMIT \cite{Garofolo1993} (i.e, ``unmatched condition") with random target azimuth and isotropic noise added at a random SNR between -6~dB and 15~dB. The speaker was placed at $0^\circ$ elevation and at a distance of either 80~cm or 300~cm chosen randomly for each signal. Reverberant speech signals for evaluation were generated using \acf{BRIR}s from \cite{Kayser2009} and were placed in isotropic noise fields for the anechoic signals. Rooms with $T_{60}$ varying from $0.3$ to $1.2$~s were used.
\vspace{-0.2cm}
\subsection{Training setup and baselines}
For the \ac{STFT} computation, an FFT length of 512, a window length of 25~ms, and a hop length of 6.25~ms were used. A sampling rate of 16~kHz was used for all signals. The following methods were used for the evaluation and comparison to the proposed binaural enhancement model.

 \textbf{\ac{BCCTN}}: This is our proposed method. The number of channels used in the \ac{MIMO} model's convolutional layers for the encoder and decoder blocks layers are $\{16, 32, 64, 128, 256, 256\}$, with a stride of 2 in the frequency and 1 in the time dimension with a kernel size of (5,1) and all the convolutions in these layers are causal. The Multihead attention block has an embedded dimension of 512 for real and imaginary blocks shown in Fig.~\ref{fig:BlockDiagtrans}, a hidden size of 128, and 32 heads. The model was implemented with Pytorch which provides native complex data support for most of the functions. The linear layer placed after the transformer block has an input and output feature size of 1024. The Pytorch model was trained using the Adam optimizer, an initial learning rate of 0.001, and a multi-step learning rate scheduler to modify the learning rate with the validation loss. The model has around 10 million parameters and was trained for 100 epochs with an additional early stopping condition of no improvement in the validation loss for three consecutive epochs. The loss functions weights $\alpha, \beta, \gamma, \kappa$, in \eqref{lossfunc}, were set to  $\{1,10,1,10\}$ respectively. These weights were chosen to equalize the difference in the scale of the respective units of the individual loss function terms where \ac{SNR} and \ac{ILD} are computed in dB, \ac{IPD} is computed in radians and \ac{STOI} is a bounded score between 0 and 1. The model was trained with the proposed loss function described in \eqref{lossfunc} and, for comparison, the model was also trained to maximize the \ac{SNR} from \eqref{snr}.


 \textbf{\ac{BSOBM}}: A binaural speech enhancement method using \ac{STOI}-optimal masks proposed in \cite{Tokala2022}. Here a feed-forward \ac{DNN} was trained to estimate a \ac{STOI}-optimal continuous-valued mask to enhance binaural signals using dynamically programmed \ac{HSWOBM} as the training target \cite{Tokala2022}. To preserve the \ac{ILD}s, a better-ear mask was computed by choosing the maximum of the two masks.  The mask is used to supply \ac{SPP} to an \ac{OM-LSA} enhancer. The model was trained and evaluated on the same dataset as the proposed model.

 \textbf{\ac{BiTasNet}}: A time-domain \ac{MIMO} \ac{CED}-based network for binaural speech separation which was introduced in \cite{Han2020}. The best-performing version of the model, the parallel encoder with mask and sum, was modified and retrained for single-speaker binaural speech enhancement. The network was trained to maximize \ac{SNR}\cite{Han2020}. The encoder and decoders in the model had a size of 128, a feature dimension of 128, kernel size of 3 and 12 layers. All other parameters were adapted from the original article and the model has a size of 9.7 million parameters. The model was trained and evaluated on the same dataset used for the proposed method.

\vspace{-0.15cm}
\section{Results and Discussion}
\label{sec:results}
The model was evaluated using 750 speech utterances from both datasets for each noisy input SNR. In total, the model was evaluated on 6000 noisy speech utterances. Improvement in the frequency weighted \ac{SegSNR} \cite{Brookes1997} was used to show the noise reduction performance of the methods. The \ac{MBSTOI} \cite{Andersen2018} score was computed to measure the objective binaural speech intelligibility of the enhanced signals. The error in \ac{ILD} and \ac{IPD} after processing were computed using equations \eqref{ild} and \eqref{ipd} respectively to evaluate the preservation of interaural cues. Tables \ref{tab_anechoic} and \ref{tab_reverb} show the results tabulated for multiple \ac{SNR}s for anechoic and reverberant speech signals respectively. For noise reduction measured by the improvement ($\Delta$) in frequency weighted \ac{SegSNR} \cite{Brookes1997}, \ac{BiTasNet} has the best performance with \ac{SegSNR} for almost all \ac{SNR}s. However, the proposed method shows comparable performance to \ac{BiTasNet} on the noise reduction task for both \ac{SNR}-optimization and the proposed loss function. The proposed loss function had better noise reduction performance compared to the model with the \ac{SNR} loss function. A possible explanation is that the addition of intelligibility and masked interaural cue terms in the loss function enables the network to better identify the active speech regions which results in better noise reduction performance. A maximum of 14~dB of \ac{SegSNR} can be observed when the signal is very noisy at -6~dB SNR. Even though the \ac{BiTasNet} has better noise reduction performance, it exhibits a lower \ac{MBSTOI} binaural intelligibility score. Informal listening tests revealed that the \ac{BiTasNet} produced more artefacts. Audio examples of all the methods can be found online \footnote{\url{https://vikastokala.github.io/bse_dcctn/
}}. The model provides an average of 0.15 to 0.25 improvement in \ac{MBSTOI} scores over the noisy speech when the SNR is below 6~dB. As the input signal's SNR improves, the noisy signals inherently have a higher \ac{MBSTOI}, and the proposed model provides a lower improvement. In cases with high input \ac{SNR}, the \ac{BSOBM}, \ac{BiTasNet} and \ac{BCCTN}-SNR methods degrade the \ac{MBSTOI} score due to processing but the proposed method and loss function do not reduce the score or deteriorate the signal at high \ac{SNR}s. The proposed model and loss function have the lowest \ac{ILD} and \ac{IPD} error for all \ac{SNR}s. The proposed model with \ac{SNR} loss function performs similarly to the proposed loss function in noise reduction but does not focus on retaining the interaural differences and the additional terms in the loss function help the network in the preservation of interaural cues better. From Table \ref{tab_reverb}, similar performance trends for reverberant signals can be observed from all the methods. A maximum of 10~dB of \ac{SegSNR} can be observed when the signal is very noisy and up to a maximum of 0.15 improvement in \ac{MBSTOI} score. The \ac{ILD} and \ac{IPD} errors are slightly higher than the anechoic condition which could be due to the effects of reverberation \cite{Han2020}.

\section{Conclusion}
\label{sec:conclusion}
In this paper, we have presented a \ac{MIMO} complex-valued convolutional transformer network for binaural speech enhancement. A novel loss function that optimizes the network for noise reduction, speech intelligibility enhancement, and interaural cue preservation is proposed. Experimental results show that the proposed method was able to significantly reduce noise and has the ability to preserve \ac{ILD} and \ac{IPD} information in the enhanced output. Furthermore, the proposed method outperforms the baselines in terms of estimated binaural speech intelligibility. Future works include adapting the model to include a remote microphone and a distributed microphone network for binaural speech enhancement.


\bibliographystyle{IEEEtran}
\bibliography{sapstrings,sapref}
%
%
%
%
%
%
%
%
%

\end{sloppy}
\end{document}